\newif\ifjournal
\journaltrue
\ifjournal
  \documentclass{aa}
  \usepackage{graphicx,txfonts,amssymb,natbib}
  \renewcommand{\d}{\mathrm{d}}
  \authorrunning{Maturi et al.}
  \titlerunning{Gravitational lensing of the CMB by galaxy clusters}
\else
  \documentclass{paper}
\fi

\newcommand{\ii}{\mathrm{i}}
\newcommand{\e}{\mathrm{e}}

\begin{document}

\title{Gravitational lensing of the CMB by galaxy clusters}
\ifjournal
\author{Matteo Maturi\inst{1,2,3}, Matthias Bartelmann\inst{2},
  Massimo Meneghetti\inst{2} and Lauro Moscardini\inst{4}}
\institute
 {$^1$ Dipartimento di Astronomia, Universit\`a di Padova, vicolo
  dell'Osservatorio 2, I-35122 Padova, Italy\\
  $^2$ Institut f\"ur Theoretische Astrophysik, Universit\"at
  Heidelberg, Tiergartenstr.~15, D--69121 Heidelberg, Germany\\
  $^3$ Max-Planck-Institut f\"ur Astrophysik, P.O.~Box 1317, D--85740
  Garching, Germany\\
  $^4$ Dipartimento di Astronomia, Universit\`a di Bologna, via
  Ranzani 1, I-40127 Bologna, Italy}
\else
\author{Matteo Maturi$^{1,2,3}$, Matthias Bartelmann$^2$, Massimo
  Meneghetti$^2$ and Lauro Moscardini$^4$\\
  $^1$ Dipartimento di Astronomia, Universit\`a di Padova, vicolo
  dell'Osservatorio 2, I-35122 Padova, Italy\\
  $^2$ Institut f\"ur Theoretische Astrophysik, Universit\"at
  Heidelberg, Tiergartenstr.~15, D--69121 Heidelberg, Germany\\
  $^3$ Max-Planck-Institut f\"ur Astrophysik, P.O.~Box 1317, D--85740
  Garching, Germany\\
  $^4$ Dipartimento di Astronomia, Universit\`a di Bologna, via
  Ranzani 1, I-40127 Bologna, Italy}
\fi

\date{Received/Accepted}

\newcommand{\abstext}
 {We adapt a non-linear filter proposed by \cite{HU01.1} for detecting
  lensing of the CMB by large-scale structures to recover
  surface-density profiles of galaxy clusters from their localised,
  weak gravitational lensing effect on CMB fields. Shifting the
  band-pass of the filter to smaller scales, and normalising it such
  as to reproduce the convergence rather than the deflection angle, we
  find that the mean density profile of a sample of 100 clusters can
  be recovered to better than 10\% from well within the scale radius
  to almost the virial radius. The kinetic Sunyaev-Zel'dovich effect
  is shown to be a negligible source of error. We test the filter
  applying it to data simulated using the characteristics of the
  \emph{Atacama Cosmology Telescope} (ACT), showing that it will be
  possible to recover mean cluster profiles outside a radius of $1'$
  corresponding to ACT's angular resolution.}

\ifjournal
  \abstract{\abstext
    }
\else
  \begin{abstract}
  \abstext
  \end{abstract}
\fi

\maketitle

\section{Introduction}

Future observations of large fractions of the sky in the sub-mm regime
are expected to detect of order $10^4$ galaxy clusters through their
thermal Sunyaev-Zel'dovich (tSZ) effect
\citep{SI00.1,BA01.3,CA02.1}. They appear projected on the CMB, which
is relatively feature-less on the angular scales typical for galaxy
clusters. The weak gravitational lensing caused by the clusters will
distort the CMB pattern in a characteristic way \citep{SE00.1}. It
appears unlikely that the lensing signal of individual clusters on the
CMB will be detectable with current experiments, but it may be
possible to extract statistical information on the cluster mass
distribution by suitably combining the signal caused by large cluster
subsamples. Future instruments like ALMA with high sensitivity and
resolution should even be able to detect lensing by few or single
clusters.

Various possibilities for detecting cluster lensing signals in the CMB
were recently discussed. \cite{HO04.1} suggested to use Wiener
filtering for estimating the unlensed temperature map, which is then
subtracted from the measured lensed map to obtain the deflection
field. \cite{VA04.1} proposed a conceptually similar technique based
on subtracting an estimate of the unlensed CMB background, obtained by
fitting a temperature gradient to the observed map.

We follow a different route, starting from a non-linear filter
suggested by \cite{HU01.1} for extracting the lensing signal of
large-scale structure from CMB observations. Since the CMB on cluster
scales can be locally approximated by a gradient, the cluster-lensed
CMB temperature pattern is randomly oriented. Aiming at a technique
which allows cluster-sized CMB images to be stacked in order to
extract statistical information on cluster profiles, we thus require a
filter which is non-linear in the signal. \citeauthor{HU01.1}'s filter
essentially measures the squared gradient of the temperature map and
thereby renders the filtered maps suitable for stacking. We modify the
filter in two ways which turn out to improve it substantially when
applied to lensing by clusters rather than large-scale structures.

The plan of the paper is as follows: We summarise CMB lensing and
introduce the filter in Sect.~2. In Sect.~3, we describe simulations
of clusters and cluster samples. In Sect.~4, we apply the filter to
these simulations, taking the kinetic Sunyaev-Zel'dovich (kSZ) effect
and instrumental noise into account. We summarise and discuss our
results in Sect.~5.

\section{Lensing of the CMB and filtering}

Lensing by a thin mass distribution on a single plane can be described
by the scalar lensing potential
\begin{equation}
  \psi(\vec\theta)=
  \frac{2}{c^2}\frac{D_\mathrm{ds}}{D_\mathrm{d}D_\mathrm{s}}
  \int\Phi(D_\mathrm{d}\vec\theta,z)\d z\;,
\label{eq:0}
\end{equation}
where $D_\mathrm{d,s,ds}$ are the angular-diameter distances from the
observer to the lens, the source, and from the lens to the source,
respectively, and $\Phi$ is the Newtonian gravitational potential of
the lensing mass distribution, which is integrated along the
line-of-sight. The deflection angle $\vec\alpha=\vec\nabla\psi$, and
the convergence $\kappa=\vec\nabla^2\psi/2$ is the scaled surface-mass
density \citep[see e.g.~][for reviews]{SC92.1,NA99.1}.

Light rays propagating into a direction $\vec\theta$ on the sky are
deflected by the angle $\vec\alpha(\vec\theta)$. Thus, the CMB
temperature $T(\vec\theta)$ \emph{observed} into direction
$\vec\theta$ is the intrinsic temperature $\tilde T$ at a slightly
shifted position,
\begin{equation}
  T(\vec\theta)=\tilde T[\vec\theta-\vec\alpha(\vec\theta)]\approx
  \tilde T(\vec\theta)-
  \vec\nabla\tilde T(\vec\theta)\cdot\vec\alpha(\vec\theta)
\label{eq:1}
\end{equation}
\citep{SE00.1}. The approximation used here is justified if the
intrinsic CMB temperature does not vary much on angular scales which
are characteristic for lensing by galaxy clusters. In fact, on the
arc-minute scales of weak cluster lensing, the CMB is almost
feature-less and can to first order be approximated by a gradient.

It is clear from Eq.~(\ref{eq:1}) that all information on the lensing
potential $\psi$ is contained in the lensed map through the
deflection-angle field
\begin{equation}
  \vec\alpha(\vec\theta)=\vec\nabla\psi(\vec\theta)=
  \int\frac{d^2\vec l}{(2\pi)^2}i\vec l\,
  \psi(\vec l)e^{i\vec l \vec\theta}\;,
\end{equation}
thus it is in principle possible to recover all the information
contained in $\psi$, like for instance the deflection-angle field
itself or the lensing convergence.

The lensing deflection (\ref{eq:1}) gives rise to a characteristic
distortion of the local CMB temperature gradients on which clusters
happen to appear projected. However, the orientation of the signal
reflects the random orientation of $\vec\nabla\tilde T$. Stacking
clusters, which we anticipate will be necessary because of the weak
signal of a single cluster, thus tends to erase the signal if
temperature maps or linear transformations thereof are used.

It is possible to estimate the lensing-induced distortion of the CMB
temperature by smoothing over the cluster and fitting a gradient to
the resulting field, or replacing the cluster area by a suitably
adapted gradient or polynomial functions \citep{HO04.1,VA04.1}. Apart
from the fact that any estimate derived from fitting has to vanish at
the field boundaries, orientation angles of $\vec\nabla\tilde T$
estimated that way will be uncertain. Stacking cluster fields after
lining them up according to their estimated temperature-gradient
orientations is therefore still likely to erase a considerable
fraction of the signal.

\cite{HU01.1} suggested a non-linear filter which extracts a scalar
measure for the lensing pattern from CMB observations. He started from
an essentially Wiener-filtered CMB temperature gradient,
\begin{equation}
  \vec G(\vec\theta)=\int\frac{\d^2\vec l}{(2\pi)^2}
  \frac{C_l}{C_l^\mathrm{tot}}\ii\vec l\,T(\vec l)\,
  \e^{\ii\vec l\vec\theta}\;,
\label{eq:2}
\end{equation}
where $C_l$ is the CMB power spectrum, and
$C_l^\mathrm{tot}=C_l+C_l^\mathrm{noise}$ is the sum of the power
spectra of signal and noise. This is then multiplied with a suitably
high-pass filtered temperature map $W(\vec\theta)$ to obtain the
product
\begin{equation}
  \tilde{\vec G}(\vec\theta)=W(\vec\theta)\vec G(\vec\theta)
\label{eq:3}
\end{equation}
of which the filtered divergence is taken in Fourier space to obtain
the ``deflection field''
\begin{equation}
  D(\vec\theta)=-\int\frac{\d^2L}{(2\pi)^2}\frac{N_L}{L}\,
  \ii\vec L\cdot\tilde{\vec G}(\vec L)\;.
\label{eq:4}
\end{equation}
\cite{HU01.1} showed that the averaged deflection field $D(\vec L)$
can be written as
\begin{equation}
  \langle D(\vec L)\rangle=L\psi(\vec L)\;,
\label{eq:5}
\end{equation}
provided $N_L$ is suitably chosen; $\psi(\vec L)$ is the Fourier
transform of the lensing potential, whose gradient is the deflection
angle, $\vec\alpha=\vec\nabla\psi$. In finding $N_L$, the first-order
expansion of (\ref{eq:1}) was used.

As we shall demonstrate later, it is important for our purposes to
change \citeauthor{HU01.1}'s original filter somewhat, modifying the
definition of $W(\vec\theta)$ and aiming at filtering for the cluster
convergence $\kappa$ rather than the deflection field
$D(\vec\theta)$. We use
\begin{equation}
  W(\vec\theta)=\int\frac{\d^2l}{(2\pi)^2}
  \frac{l}{C_l^\mathrm{tot}}\,T(\vec l)\,\e^{\ii\vec l\vec\theta}\;,
\label{eq:6}
\end{equation}
which has an additional factor $l$ in the numerator compared to
\citeauthor{HU01.1}'s definition. This makes the filter more sensitive
to smaller scales, which is crucial for detecting cluster lensing, as
will become obvious below.

We further introduce
\begin{equation}
  K(\vec\theta)=-\int\frac{\d^2L}{(2\pi)^2}\frac{N_L'}{L}\,
  \ii\vec L\cdot\tilde{\vec G}(\vec L)
\label{eq:7}
\end{equation}
and require that $K$ on average reproduce the convergence, $\langle
K(\vec\theta)\rangle=\kappa$. Using $L^2\psi(\vec L)=2\kappa(\vec L)$,
a straightforward calculation shows that this can be achieved choosing
\begin{equation}
  (N_L')^{-1}=\frac{1}{L^3}\int\frac{\d^2\vec l_1}{(2\pi)^2}
  \frac{(\vec l\vec l_1l_2C_{l_1}+\vec l\vec l_2l_1C_{l_2})
        (\vec l\vec l_1C_{l_1}+\vec l\vec l_2C_{l_2})}
  {C^\mathrm{tot}_{l_1}C^\mathrm{tot}_{l_2}}\;,
\label{eq:8}
\end{equation}
with $\vec l_2:=\vec L-\vec l_1$. Aiming at the convergence, this
normalisation has an additional factor $L^2$ compared to the
normalisation required for filtering for the potential. Again, this
modification turns out to be crucial for the success of filtering
cluster signals. The convergence falls off more steeply than the
deflection field, which greatly simplifies the application of
fast-Fourier methods.

\section{Simulations\label{sec:simulation}}

We now describe simulations of filtering the weak cluster-lensing
signal from CMB observations. For the cosmological background model,
we adopted the standard, spatially-flat $\Lambda$CDM cosmology with
present density contributions from dark matter, baryons and
cosmological constant of $\Omega_\mathrm{DM}=0.276$,
$\Omega_\mathrm{B}=0.024$, and $\Omega_\Lambda=0.7$, respectively. The
Hubble constant was set to $H_0=100\,h\,\mathrm{km\,s^{-1}\,Mpc^{-1}}$
with $h=0.7$.

Using CMBEASY \citep{DO03.3}, we computed the CMB power spectrum to
the multipole order $\ell=30000$, assuming a reionisation fraction of
$0.1$ and a reionisation redshift of $6.2$, and adopting a present
helium abundance of $0.24$. We used this spectrum to simulate two sets
of CMB maps as Gaussian random fields. The first set has a high
resolution of $4096^2$ pixels and a field side-length of $320'$. We
will use it in Sect.~\ref{sec:ideal_case} for evaluating the
estimation of the cluster convergence profile. We use the central
quarter of these fields, i.e.~maps with $2048^2$ pixels and side
lengths of $160'$. The second set consists of lower-resolution maps
with side lengths of $280'$ and $1024^2$ pixels, of which again only
the central quarters are retained. These maps will be used for
simulating observations with the ACT telescope in
Sect.~\ref{sec:ACT}. The field sizes $S$ ensure adequate sampling of
the CMB multipoles, and are large enough to allow multiplication of
the data with the window function
\begin{equation}
  F(\vec\theta)=\left\{\begin{array}{ll}
    1 & (S/4\le\theta\le3/4S)\\
    \frac{1}{2}\left[1-\cos\left(\frac{4\pi\theta}{S}\right)\right]
    & (\hbox{elsewhere}) \\
  \end{array}\right.
\end{equation}
which we apply before filtering to impose periodic boundary conditions
for fast-Fourier transforms, leaving the central part unchanged.

\begin{figure}[ht]
  \includegraphics[width=\hsize]{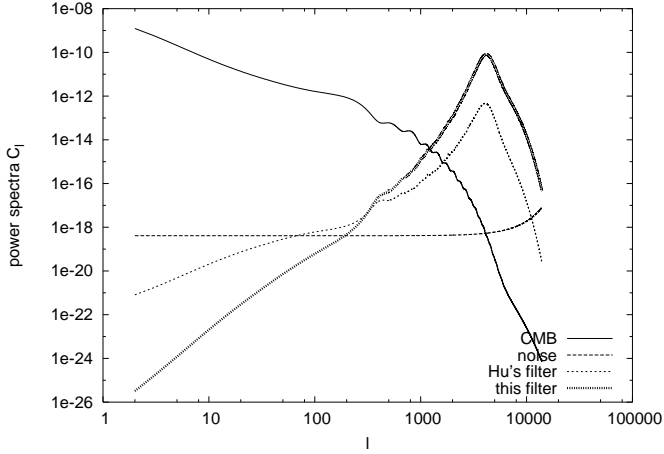}
\caption{Power spectra of the CMB (solid line), the instrumental noise
  (long-dashed line), \citeauthor{HU01.1}'s original filter function
  (short-dashed line) and our modified filter defined in
  Eq.~(\ref{eq:6}, heavy dotted line). The noise was modelled such as
  to reproduce the 225~GHz ACT channel. It is obvious that $W$ is a
  band-pass filter centred on the minimum of the combined CMB and
  instrumental noise. Our modified version of $W$ suppresses more
  efficiently the large-scale CMB noise and has much improved power on
  smaller scales in the cluster regime.}
\label{fig:spettri}
\end{figure}

While the tSZ effect can be ignored working at 217~GHz, the kSZ effect
needs to be taken into account. In order to simulate it, we assume
that the intracluster gas is isothermal and has a radial $\beta$
profile with $\beta=1$ (i.e.~a King profile). The Compton-parameter
profile is then
\begin{equation}
  y(\theta)=y_0\left[
    1+\left(\frac{\theta}{\theta_\mathrm{c}}\right)^2
  \right]^{-1}\;,
\label{eqn:Compton}
\end{equation}
where $\theta_\mathrm{c}$ is the core radius, and $y_0$ is determined
by the integrated Compton parameter $Y$. For an isothermal cluster at
an angular-diameter distance of $D(z)$ containing $N_\mathrm{e}$
electrons with temperature $T_\mathrm{e}$, it is
\begin{equation}
  Y=\frac{kT_\mathrm{e}}{m_\mathrm{e}c^2}
  \frac{\sigma_\mathrm{T}}{D^2(z)}N_\mathrm{e}\;,
\label{eqn:Compton_integrato}
\end{equation}
where $k$ is Boltzmann's constant, $c$ is the speed of light,
$m_\mathrm{e}$ is the electron rest mass, and $\sigma_\mathrm{T}$ is
the Thomson scattering cross section. If the line-of-sight velocity of
the cluster is $v_\parallel$, its kSZ effect gives rise to the
temperature fluctuation
\begin{equation}
  \frac{\Delta T}{T}=-\frac{m_\mathrm{e}c}{kT_\mathrm{e}}\,
  v_\parallel\,y(\theta)\;,
\label{eqn:DTcinelatico}
\end{equation}
where we have assumed isothermal clusters without significant internal
gas motion.

We adopt the NFW density profile \citep{NA97.1} for the dark-matter
distribution of the clusters. An analytic equation for its deflection
angle is given in \cite{BA96.1} and was used to compute the lensed CMB
temperature maps using a linear expansion (\ref{eq:1}) of the lens
equation.

Instrumental noise and resolution effects were included adding to the
lensed temperature map a noise contribution computed as a Gaussian
random field with the power spectrum
\begin{equation}
  C_l^\mathrm{noise}=w^{-1}\exp\left[
  \frac{l(l+1)\,\mathrm{FWHM}^2}{8\ln2}
  \right]\;,
\end{equation}
where $w^{-1}:=(\Delta T/T\,\mathrm{FWHM})^2$ \citep{KN95.1}. The
lensed temperature map with the noise added was finally convolved with
a Gaussian kernel with the same FWHM. The instrumental effect was
applied only in simulating the 225~GHz channel of the \emph{Atacama
Cosmology Telescope} (ACT, \citealt{KO04.1}) as explained in
Sect.~\ref{sec:ACT}.

We show in Fig.~\ref{fig:spettri} a comparison between the power
spectra of the CMB (solid line), the instrumental ACT noise
(long-dashed line), the filter $W$ as given in Eq.~(\ref{eq:6}, heavy
dotted line) and its original version suggested by
\citeauthor{HU01.1} (long-dashed line). The amplitudes were
arbitrarily rescaled in order to improve the graphical representation.

\begin{figure}[ht]
  \includegraphics[width=\hsize]{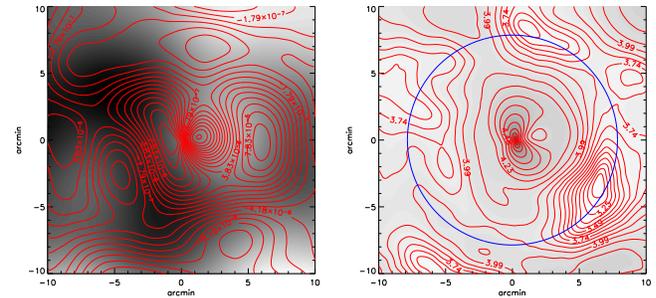}
\caption{Left panel: Simulated CMB temperature map with overlaid
  iso-contours of the temperature distortion caused by the lensing
  effect of a single galaxy cluster. Right panel: Convergence
  estimated with the non-linear filter described in the text. The
  background grey-scale map has a linear scale, while the iso-contours
  are spaced logarithmically. The circle represents the virial radius
  of the cluster. The maps show results for an ideal instrument
  without noise and a resolution of $4.68$~arc sec. Note that the
  reconstructed convergence shows a spurious elongation perpendicular
  to the local CMB gradient, which arises because the lensing effect
  on the CMB vanishes in this direction.}
\label{fig:un_cluster}
\end{figure}

\section{Data analysis: Application of the filter}

We demonstrate in this Section the filter's capability to recover the
convergence by first applying it to a single cluster in the ideal case
without any instrumental noise (Subsect.~\ref{sec:ideal_case}) and
then including the kSZ effect (Subsect.~\ref{sec:SZ}).

In Subsect.~\ref{sec:ACT}, we investigate whether and how it will be
possible with the data from the 225-GHz \emph{ACT} channel to obtain
average cluster convergence profiles. The assumed observing strategy
is to detect clusters through their tSZ signal and then to suitably
stack CMB maps surrounding them in order to enhance the
signal-to-noise ratio. We shall assume that a total area of
$200$~square degrees will be covered with ACT.

\subsection{Convergence estimation for a single
  cluster\label{sec:ideal_case}}

For studying the filter's capability to recover the convergence of a
single cluster, we applied our filter to a CMB map with $2048^2$
pixels simulated as described in Sect.~\ref{sec:simulation}. This CMB
map was lensed with an axially-symmetric galaxy cluster model with NFW
density profile and with a mass of $10^{15}\,h^{-1}\,M_\odot$ placed
at redshift $0.3$. No instrumental noise was added yet.

The left panel of Fig.~\ref{fig:un_cluster} shows the CMB temperature
map with overlaid iso-contours of the temperature distortion pattern
caused by the lensing effect of the cluster. The characteristic dipole
pattern is clearly visible. The right panel of the same figure shows
the estimated cluster convergence. The grey-scale map has a linear
scale, while the iso-contours are logarithmically spaced. The circle
illustrates the virial radius of the cluster.

\begin{figure}[ht]
  \includegraphics[width=\hsize]{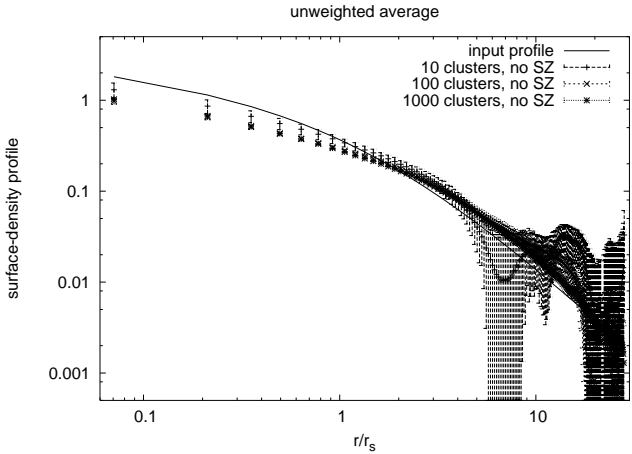}
\caption{The convergence profile of the input cluster is plotted as
  the solid line. The dots with error bars show the reconstructed mean
  convergence profile determined from 10, 100, and 1000 replications
  of the same cluster on different realisations of the CMB. The error
  bars were determined by bootstrapping the cluster sample. While the
  input profile is well reproduced outside a few scale radii, the
  reconstructed profile is too flat in the core.}
\label{fig:1}
\end{figure}

Figure~\ref{fig:1} illustrates the cluster profile reconstruction
after a straightforward application of our filter. The input cluster,
whose profile is plotted as the solid curve, is placed in front of 10,
100, and 1000 different realisations of the CMB. The filter is
applied, and the radial profile of the convergence estimate
$K(\vec\theta)$ is determined and averaged across the cluster
sample. Outside a few scale radii, the reconstructed profile flattens
because the clusters disappear in the CMB. While approaching the input
profile outside a few scale radii until the noise begins dominating,
the reconstructed profile is evidently too flat in the core. This can
be overcome by introducing a suitably weighted average, as we shall
now describe.

Note that the reconstructed convergence shows a spurious elongation
perpendicular to the local CMB gradient, which arises because lensing
of the CMB by a cluster has no effect in this direction
(cf.~Eq.~\ref{eq:1}). This degeneracy is thus characteristic for
lensing of the CMB temperature anisotropy rather than the filtering
technique applied. It could be broken only with some additional
information, like lensing of the polarisation anisotropy.

This suggests to determine the convergence profile not by a straight
average in azimuthal bins, but after multiplying the observed map with
a weight map
\begin{equation}
  P(\vec\theta)=|\vec G(\vec\theta)|^2\cos^2\xi(\vec\theta)\;,
\label{eqn:w_ave}
\end{equation}
where $\xi(\vec\theta)$ is the angle between the CMB gradient and the
deflection field at the position $\vec\theta$, under the assumption
that it has radial symmetry. This is exact for our axially-symmetric
analytical cluster model, but it can also be applied to the general
case of asymmetric clusters because the deviation of the
deflection-angle direction from the radial symmetry is small even for
substructured clusters. Assuming therefore that the deflection-angle
points towards the field centre, we have
\begin{equation}
  \cos\xi(\vec\theta)=
  -\frac{\vec\nabla T(\vec\theta)}{|\vec\nabla T(\vec\theta)|}\cdot
  \vec e_r=-\frac{\vec G(\vec\theta)}{|\vec G(\vec\theta)|}\cdot
  \vec e_r
\label{eq:cosine}
\end{equation}
where $\vec e_r$ is the radial unit vector. Note that we have used the
Wiener-filtered temperature gradient $\vec G(\vec\theta)$ introduced
in Eq.~\ref{eq:2} which has its noise component minimised.

Using $P(\vec\theta)$, we determine radial weight profiles
$P_i(\theta)$ for each cluster $i$ by averaging azimuthally,
\begin{equation}
  P_i(\theta)=\theta\int_0^{2\pi}\d\varphi P_i(\vec\theta)\;,
\label{eq:weightp}
\end{equation}
and then determine the averaged convergence profile across a cluster
sample,
\begin{equation}
  \langle\kappa\rangle(\theta)=
  \frac{\sum_i\kappa_i(\theta)P_i(\theta)}{\sum_iP_i(\theta)}\;.
\label{eq:weighta}
\end{equation}

\begin{figure}[ht]
  \includegraphics[width=\hsize]{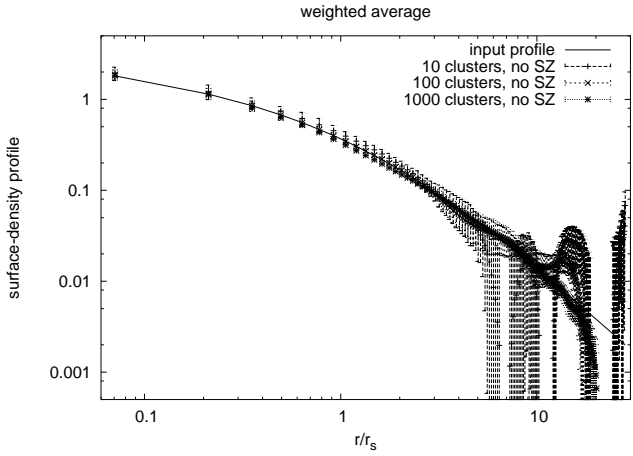}
\caption{True and reconstructed cluster convergence profiles are shown
  as in Fig.~\ref{fig:1}, but now using the weighted profile average
  described in the text. The profile is now accurately reproduced well
  into the cluster core. Averaging over 100 clusters, the profile is
  detected out to approximately ten scale radii, or roughly twice the
  virial radius.}
\label{fig:2}
\end{figure}

As in Fig.~\ref{fig:1} for the unweighted average, Fig.~\ref{fig:2}
shows the recovered cluster convergence profiles obtained after
weighting the average as described in (\ref{eq:weighta}). Well into
the cluster core, the profile is now accurately reproduced. Averaging
over a sample of 100 clusters, the profile is detected out to
approximately ten scale radii, roughly corresponding to two virial
radii. In view of the persisting uncertainty of central cluster
density profiles, it is promising that the slope of the core profile
is accurately reproduced even from ten clusters only.

Clusters which happen to be superposed on extrema or saddle points of
the CMB temperature contribute little or nothing to the average
lensing signal. The weighting scheme introduced here takes this
automatically into account by reducing their statistical weight
according to the signal they contribute.

The power of this filtering method is evident and opens a new
technique for analysing data from the next-generation, high-resolution
sub-millimetre observatories like ALMA\footnote{ALMA home page,
http://www.eso.org/projects/alma/}.

\begin{figure}[ht]
  \includegraphics[width=\hsize]{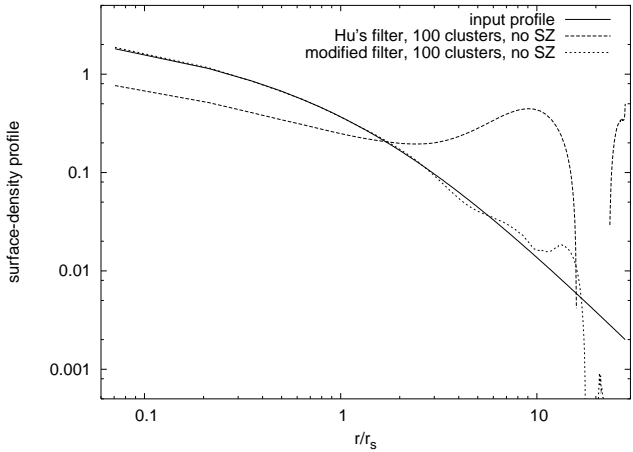}
\caption{Cluster convergence profiles as recovered by Hu's
  (\citeyear{HU01.1}) original filter (long-dashed curve) and our
  modification of it (short-dashed curve). The input profile is shown
  as the solid curve. Both reconstructed profiles are weighted
  averages over 100 clusters.}
\label{fig:5}
\end{figure}

\cite{HU01.1} originally proposed his filter for recovering the
deflection-field of the large-scale structure. In order to filter for
clusters, our modification (\ref{eq:6}) of the weight map
$W(\vec\theta)$ is crucial because introducing the factor $l$
increases the filter's sensitivity on small scales. As
Fig.~\ref{fig:5} shows, the resulting effect is dramatic. While Hu's
filter certainly detects the cluster, but fails in recovering its
profile, our modified filter faithfully reproduces the input profile
over almost two orders of magnitude in the radius.

\subsection{Adding the kinetic Sunyaev-Zel'dovich
  effect\label{sec:SZ}}

As described, this non-linear filtering procedure was optimised for
filtering out the uncorrelated Gaussian noise, like the CMB itself,
instrumental noise and residuals from foreground subtraction, but it
turns out to be also quite powerful in filtering out the contribution
from the kSZ effect. The latter has a characteristic which could lead
to future further improvements, as we shall clarify below.

\begin{figure}[ht]
  \includegraphics[width=\hsize]{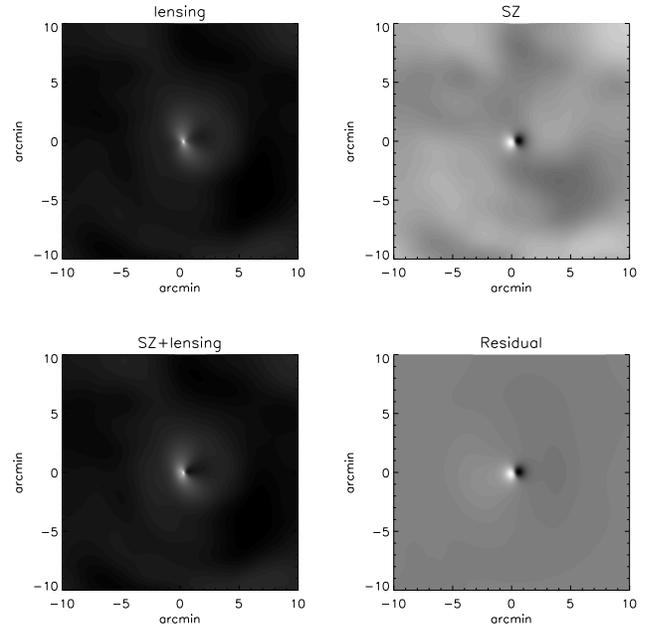}
\caption{Processing of the kSZ effect: Results are shown for the
  cluster convergence reconstructed from a CMB field, to which only
  the lensing effect (top-left panel), only the kSZ effect (top-right
  panel), and the combination of both (bottom-left panel) was
  applied. The difference between the convergences recovered from the
  lensed CMB map and from the lensed CMB map with kSZ effect is shown
  in the bottom-right panel. The kSZ effect is processed like lensing
  by a hypothetical object with positive and negative convergence. The
  figure shows that the kSZ effect is well suppressed by the filter.}
\label{fig:un_SZ}
\end{figure}

In order to understand how the filtering procedure processes the kSZ
effect, we analysed a simulated map consisting only of the CMB and the
kSZ effect. The result is shown in the top-right panel of
Fig.~\ref{fig:un_SZ}. The top-left panel shows the same realisation of
the CMB lensed by a cluster, the bottom-left panel combines both top
panels, and the difference between the bottom-right and top-right
panels is displayed in the bottom-right panel. The kSZ effect
introduces an artefact in the recovered convergence field which has a
maximum amplitude of only one tenth of the lensing signal and is
characterised by a dipolar structure aligned with the CMB gradient,
along which the lensing effect is also the strongest. Azimuthally
averaging the recovered convergence, and stacking many cluster fields
with random intrinsic orientations of the CMB gradient, further
suppresses the kSZ signal in the lensing reconstruction.

\begin{figure}[ht]
  \includegraphics[width=\hsize]{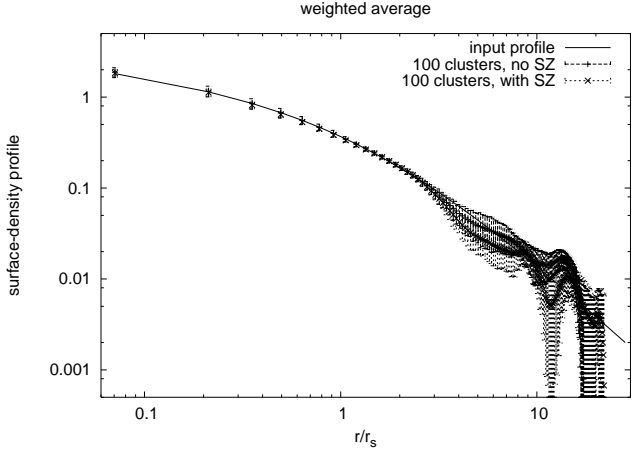}
\caption{Comparison of the recovered convergence profiles ignoring and
  including the kSZ effect (points with long- and short-dashed error
  bars, respectively). The solid curve shows the input cluster
  profile. Due to the various reasons detailed in the text, the kSZ
  effect leaves the lensing reconstruction essentially unaffected.}
\label{fig:3}
\end{figure}

The dipolar pattern in the recovered convergence is caused by the fact
that the filter interprets the entire secondary anisotropy as a
lensing signal, and mimics it with a hypothetical lens with two
adjacent positive and negative convergence peaks. Such a lens has a
deflection-angle field reminiscent of an electric dipole, leading to a
temperature signature identical to that of the kSZ effect. This
peculiarity of the characteristic kSZ pattern could be used to further
improve the filter including criteria for simultaneously minimising
the kSZ effect.

Figure~\ref{fig:3} illustrates the recovery of the cluster convergence
profile in presence of the kSZ effect. Both curves with error bars
were obtained placing the same cluster in front of 100 CMB
realisations each and determining the weighted convergence profile as
before. Evidently, the kSZ effect does not affect the lensing
reconstruction in any significant way.

\begin{figure}[ht]
  \includegraphics[width=\hsize]{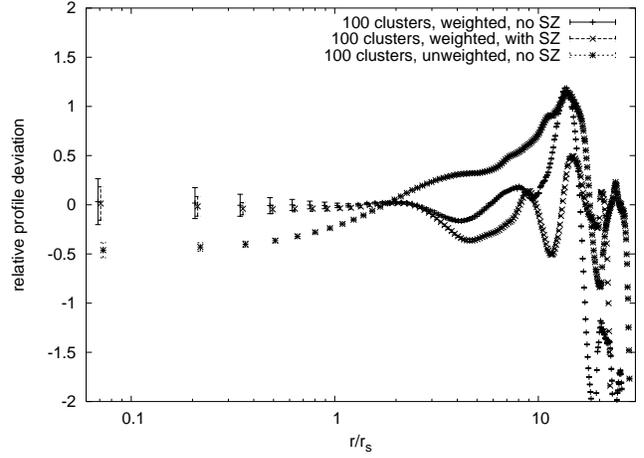}
\caption{Relative deviation of the recovered cluster convergence
  profile averaging over three samples of 100 clusters each. The
  results were obtained weighting the average, ignoring or including
  the kSZ effect (solid and long-dashed error bars, respectively), and
  ignoring the kSZ effect and not weighting the average (short-dashed
  error bars). Weighting the average, the average profile is recovered
  to better than 10\% near the scale radius.}
\label{fig:4}
\end{figure}

Summarising this section, we show in Fig.~\ref{fig:4} the relative
deviation of the reconstructed cluster convergence profiles for three
samples of 100 clusters each, without and with weighting the average,
and without and with the kSZ effect included. Employing the weighted
average, the recovered convergence is accurate to better than ten per
cent near the scale radius, irrespective of the kSZ effect.

\subsection{Example: Application to the cluster sample expected from
  ACT\label{sec:ACT}}

For this Section, we simulated Gaussian CMB maps, including the
lensing and the kSZ signal of clusters detectable through their tSZ
effect, and also including the instrumental effects as explained in
Sect.~\ref{sec:simulation}. We set the noise level to
$6\,\mu\mathrm{K}$, and convolved the maps with a Gaussian beam with
$1'$ FWHM in order to investigate the expectations for estimating the
average cluster convergence profile using data from the 225-GHz
\emph{ACT} channel.

\begin{figure}[ht]
  \includegraphics[width=\hsize]{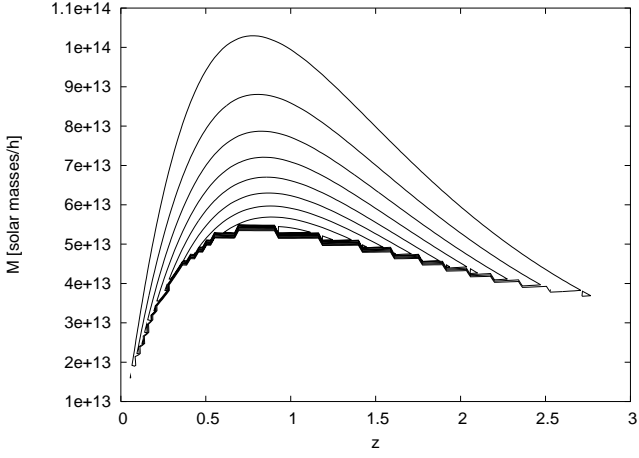}
\caption{The distribution of the galaxy-cluster sample used for this
  study is shown in the mass-redshift plane. The detection limit
  imposed by the \emph{ACT} instrumentation is well visible as the
  sharp lower cut-off. The contours show the number density of
  detectable clusters per unit mass and redshift. They range from
  $(10^{-10}-5\times10^{-10})\,hM_\odot^{-1}$ spaced by
  $5\times10^{-11}\,hM_\odot^{-1}$.}
\label{fig:cluster_distrib}
\end{figure}

We model the distribution of galaxy clusters in mass and redshift
according to the mass function proposed by \cite{SH99.1,SH02.1}. Since
galaxy clusters are detected through their tSZ effect at millimetre
and sub-millimetre wavelengths, we truncate the cluster distribution
where the solid-angle integrated Compton parameter $Y$ falls below a
threshold
\begin{equation}
  Y_\mathrm{min}=\frac{\Delta T}{T}\frac{\delta \Omega}{2}\;,
\end{equation}
where $\Delta T/T$ is the instrumental sensitivity expressed in
relative antenna temperature fluctuations, and $\delta\Omega$ is the
effective solid-angle of the beam \citep{BA01.3}. We choose
$Y_\mathrm{min}=6.6\times10^{-6}$ to approximate the sensitivity of
the 225~GHz channel on \emph{ACT} of $\Delta T/T=6\,\mu K$ and
an angular resolution of $1'$. We model the cluster distribution in
the mass range from $10^{13}\,h^{-1}\,M_\odot$ to
$10^{15}\,h^{-1}\,M_\odot$ between redshifts $0.01$ and $3$, obtaining
in total $\approx25000$ clusters. We assume that any foreground
contamination can be removed with sufficient accuracy on a survey area
of 200~square degrees. It is illustrated in
Fig.~\ref{fig:cluster_distrib}.

In order to simulate the measurement error in the cluster positions on
the sky, we randomly perturb their positions about the field centres
by an amount drawn from a Gaussian distribution with a FWHM of one
tenth of the angular resolution.

We apply the described technique to sets of CMB maps lensed by a
cluster and superposed with the kSZ effect and instrumental noise,
then compute the final weighted average convergence with the weight
described in Eq.~(\ref{eqn:w_ave}). Because of the poor resolution of
cluster cores, the weighting scheme does not noticeably improve the
results in this example, thus we show the unweighted results only.

\begin{figure}[ht]
  \includegraphics[width=\hsize]{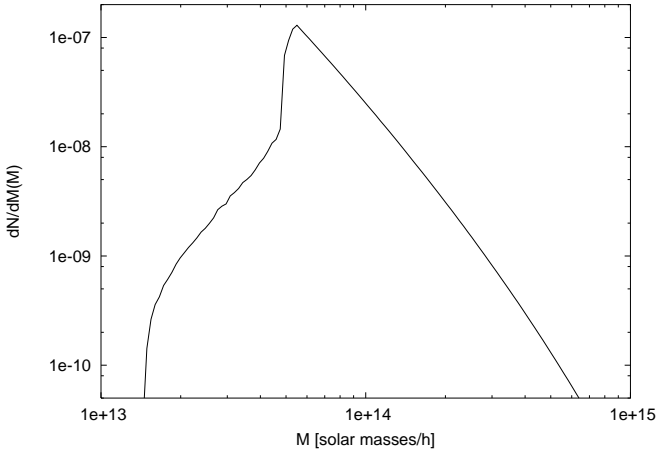}
\caption{Mass distribution of the clusters detectable for ACT. The
  lower mass cut-off is defined by beam smoothing and the detector
  sensitivity, the steep decrease towards high masses is due to the
  exponentially decreasing cluster mass function. The distribution
  peaks at $M\sim5\times10^{13}\,h^{-1}M_\odot$ and falls by one order
  of magnitude within $M\sim4\times10^{13}\,h^{-1}M_\odot$ and
  $M\sim1.2\times10^{14}\,h^{-1}M_\odot$.}
\label{fig:mass_distrib}
\end{figure}

\begin{figure}[t]
  \includegraphics[width=\hsize]{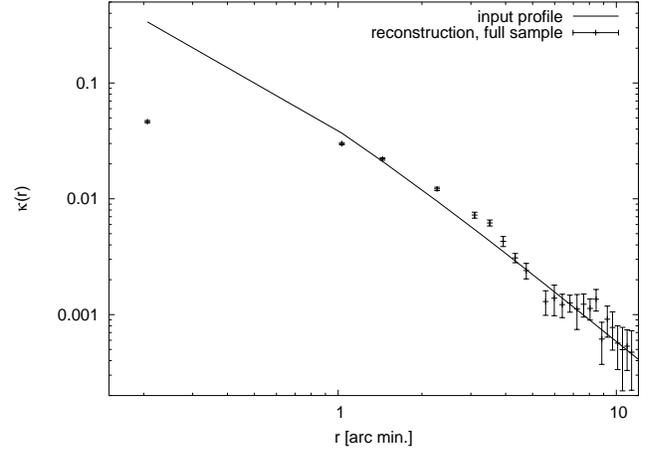}
\caption{Unweighted average estimated convergence obtained for the
  full simulated ACT cluster sample.}
\label{fig:act_result}
\end{figure}

The sharp lower cut-off in the mass shown in
Fig.~\ref{fig:cluster_distrib}, and the exponential upper cut-off due
to the steep cluster mass function, leads to a very sharp peak in the
distribution of clusters detectable for ACT
(cf.~Fig.~\ref{fig:mass_distrib}). Also, the redshift distribution of
the ACT cluster sample peaks near redshift unity, where the
angular-diameter distance is almost independent of redshift. Averaging
cluster profiles across the entire ACT sample is thus not expected to
smooth the resulting profile by more than the beam smoothing. The
result is shown in Fig.~\ref{fig:act_result}.

\begin{figure}[ht]
  \includegraphics[width=\hsize]{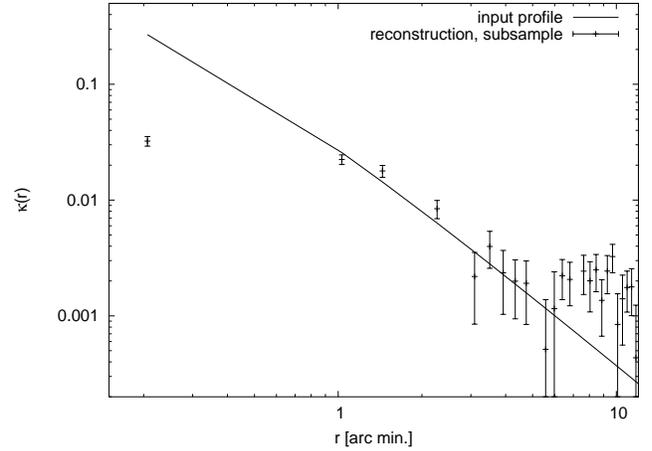}
\caption{Unweighted mean convergence profile of a subsample of the
  clusters expected to be detected by ACT. The subsample was defined
  by requiring that the integrated Compton-$Y$ parameter falls within
  $7\times10^{-6}\le Y\le7.5\times10^{-6}$.}
\label{fig:subsample}
\end{figure}

Lacking information other than that derived from the tSZ effect,
subsamples may be defined by imposing thresholds on the integrated
Compton-$Y$ parameter. As an example, we show in
Fig.~\ref{fig:subsample} the unweighted mean profile of those clusters
detected by ACT which have $7\times10^{-6}\le
Y\le7.5\times10^{-6}$. While the sample is now much better defined,
the error bars increase considerably because of the much reduced
number of clusters (2300 instead of 25000).

The instrumental resolution of $1'$ is not small compared to the
cluster size. Thus, most information is contained within a few
resolution elements. This is evident in the reconstructed profiles of
Figs.~\ref{fig:act_result} and \ref{fig:subsample}, where the measured
points fall below the input profile within $1'$, but well on the curve
at larger radii. This implies that cluster mass estimates will suffer
from the low resolution. The mean cluster mass determined from the
mean convergence profile is typically a factor of two below the true
mass of the input clusters. The lensing convergence can be converted
to mass assuming a source redshift of $1100$, and adopting the mean
redshift of the ACT clusters for the lens redshift.

\section{Conclusions}

Starting from a non-linear filter proposed by \cite{HU01.1} for
recovering the deflection-angle field of the large-scale structure
from CMB temperature fluctuations, we have constructed a non-linear
filter for extracting cluster-lensing signals from CMB maps. On the
angular scales of galaxy clusters, the CMB is almost feature-less and
can approximately be represented as a temperature gradient. Cluster
lensing imprints a characteristic pattern on that gradient, which can
be filtered for.

Since the signal obtained from a single cluster field will be weak,
current instrumentation requires stacking many cluster fields in order
to enhance the signal-to-noise ratio. Future instruments like ALMA may
allow the lensing signal of individual clusters to be detected.
Linear filtering would produce a signal aligned with the CMB gradient,
so that the signal would be removed when averaging over cluster
samples. Although other solutions are possible, this argues for a
non-linear filter which effectively squares the signal.

\cite{HU01.1} suggested to first take the Wiener-filtered temperature
gradient, high-pass filter it in order to remove large-scale noise,
take the divergence of the filtered gradient and normalise it such as
to be proportional to the deflection-angle field. In order to detect
small-scale cluster-lensing signals, we suggest to modify this
procedure by shifting the high-pass filter to even higher frequencies,
and to normalise the procedure such that the lensing surface-mass
density, or convergence, will be recovered instead of the
deflection-angle field. This has the further advantage of decaying
much more rapidly away from cluster cores, which helps in applying
fast-Fourier techniques.

We tested this non-linear filter under various assumptions. First, we
showed that cluster convergence profiles are reasonably recovered from
stacked cluster fields. Azimuthally averaging cluster fields, and
averaging the resulting profiles, reproduces the input cluster
convergence profile quite well, but the recovered profiles are too
shallow in the core.

As the reason for that deviation, we identified the fact that the
lensing signal is weak or absent in directions approximately
perpendicular to the CMB temperature gradient. We thus introduced a
weighting scheme for each individual cluster field which quantifies
the alignment of the lensing deflection with the CMB gradient, and
down-weights the signal in proportion to the cosine of the angle
between the lensing deflection and the temperature
gradient. Directions perpendicular to the temperature gradient and
flat areas in the temperature map, along which and where the lensing
signal vanishes, are thus effectively removed. Weighting the average
of the recovered cluster profiles in that way substantially improves
the agreement with the input profile. We found that an average over
100 cluster fields excellently recovers the mean cluster profile out
to several scale radii, and detects it significantly beyond the virial
cluster radius.

This filter is best applied on data taken at frequencies near 217~GHz
in order to suppress noise introduced by the tSZ effect. The kSZ
effect, however, cannot be avoided in that way. From the point of view
of gravitational lensing, the kSZ effect can be mimicked by a dipolar
lens consisting of a positive and a negative mass
contribution. Filtering lensed CMB data including the kSZ effect thus
adds a further source of noise to the recovered density
profile. Including that in our simulations, we found however that the
kSZ effect affects the results by a negligibly small amount. The three
main reasons for that are that the specific dipolar signature of the
kSZ effect tends to average away by stacking cluster fields, that mean
radial cluster velocities also average to zero across a cluster
sample, and that the spurious lensing signal produced by the kSZ
effect is approximately one order of magnitude below the true lensing
signal.

Finally, we tested what can be expected from filtering the data to be
obtained from the \emph{Atacama Cosmology Telescope} (ACT) for cluster
lensing. Given ACT's angular resolution of $1'$, cluster lensing
signals can only be detected in the innermost resolution elements
surrounding the cluster cores. It will thus only be possible to
recover mean cluster convergence profiles outside of $1'$, and mean
cluster masses will be underestimated. Those simulations took the
instrumental noise expected for ACT's 225-GHz detector into account.

These results are particularly encouraging in view of upcoming
wide-field sub-mm surveys to be carried out with telescopes like APEX
and ALMA. It appears realistic that non-linear filtering will allow
mean cluster convergence profiles to be reliably recovered from their
gravitational-lensing signature.

\acknowledgements{This work was supported in part by an EARA
  fellowship and the MIUR-COFIN 2001.}

\end{document}